\documentclass[aps,prd,groupedaddress,nofootinbib,letterpaper]{revtex4}

\usepackage{amsmath}
\usepackage{amsfonts}
\usepackage{amssymb}
\usepackage{amsthm}

\usepackage{braket,mathtools}
\usepackage{mathrsfs}
\usepackage{stmaryrd}

\usepackage{hyperref}
\usepackage{color}

\newcommand{\cale}{{\cal E}}
\newcommand{\cali}{{\cal I}}
\newcommand{\calb}{{\cal B}}

\newcommand{\beq}{\begin{equation}}
\newcommand{\eeq}{\end{equation}}
\newcommand{\bea}{\begin{eqnarray}}
\newcommand{\eea}{\end{eqnarray}}

\newcommand{\vx}{{\vec x}}

\begin{document}
\begin{titlepage}

\title{Generalized asymptotics for gauge fields}

\author{Steven B. Giddings}
\email{giddings@ucsb.edu}
\affiliation{Department of Physics, University of California, Santa Barbara, CA 93106}
\affiliation{CERN, Theory Department,
1 Esplande des Particules, Geneva 23, CH-1211, Switzerland}

\begin{abstract}
An interesting question is to characterize the general class of allowed boundary conditions for gauge theories, including gravity, at spatial and null infinity.  This has played a role in discussions of soft charges, where antipodal symmetry has typically been assumed.  However, the existence of electric and gravitational line operators, arising from gauge-invariant dressed observables, for example associated to axial or Fefferman-Graham like gauges, indicates the existence of non-antipodally symmetric initial data.  This note studies aspects of the solutions corresponding to such non-symmetric initial data.  The explicit evolution can be found, via a Green function, and bounds can be given on the asymptotic behavior of such solutions, evading arguments for singular behavior.
Likewise, objections to such solutions based on infinite symplectic form are also avoided, although these solutions may be superselected.  Soft charge conservation laws, and their modification, are briefly examined for such solutions.  This discussion strengthens (though is not necessary for) arguments that soft charges characterize gauge field degrees of freedom, but not necessarily the degrees of freedom associated to the matter sourcing the field.

\end{abstract}

\maketitle

\end{titlepage}

\section{Introduction}

An interesting and important question in the study of gauge theories, including gravity, is that of the general allowed boundary conditions for the gauge fields at infinity -- either spatial or null.  This is directly related to the question of the allowed solutions, hence degrees of freedom, of the field.  Discussion of this problem has particularly factored into discussions of soft charges in electromagnetism (EM) and gravity,\footnote{For a nice review, see \cite{astrorev}.} which have also focused on the question of what information is characterized by the soft charges, and in particular on the question of whether soft charges encode information of matter inside a black hole\cite{Hawk,HPS1,HHPS}.  

Certain important solutions -- such as the Lienard-Wiechert solution of EM -- are characterized by an antipodal symmetry, under reflection of the angular $S^2$ at spatial infinity, and a temporal reflection.  Discussions of soft charges, and their conservation, have typically assumed the presence of such a symmetry more generally.  There have also been attempts to prove the necessity of such symmetry, either as a consequence of regularity at infinity\cite{HeTrEM,Prab,Esma,HeTrHD}, or of the need for a finite symplectic form.  

It is important to understand if more general boundary conditions, and thus more general soft charge configurations, are allowed in a complete description of the physics.  Quoting \cite{HeTrEM}, ``it is always a good policy to devise boundary conditions as flexible as possible."
And in fact, construction of dressed operators, corresponding to gauge invariant observables, either in EM\cite{Dirac1955} or in gravity \cite{DoGi1}\footnote{For previous related constructions see \cite{Heem,KaLigrav}.} strongly suggest that antipodal symmetry is {\it not} a general feature of field configurations.  For example, a simple example of an operator dressed by a gravitational line arises from considering diffeomorphism-invariant operators associated to Fefferman-Graham gauge in AdS\cite{GiKi}.  These operators create a gravitational field with field lines narrowly concentrated in a particular direction, and clearly violate antipodal symmetry.  
This construction extends to, and in fact was first given with\cite{DoGi1}, the case of asymptotically Minkowski boundary conditions.
The corresponding field differs from the more symmetric ``Coulomb" (or linearized Schwarzschild) field by a pure radiation (sourceless) field.

This paper will investigate aspects of such configurations, and will focus on EM, although generalization to gravity is expected to be straightforward, based on work in \cite{DoGi1,QGQFA}.  It is first shown that such non-antipodally symmetric initial data exists,\footnote{For similar configurations in non-abelian Yang Mills, see \cite{Kosy}.} and differs from a Coulomb field by a finite-energy radiation field, which is therefore expected to disperse to infinity.  An explicit formula can be found for the corresponding solution, using the Green function for wave propagation.  One expects this solution to be finite, and finiteness of the asymptotic behavior is confirmed by investigating bounds on this solution.  Arguments for singular evolution of non-antipodal data\cite{CaEy,HeTrEM,Prab,Esma,HeTrHD} are reexamined, and found to not imply singularities in the usual electromagnetic fields.  The question of the symplectic form is also briefly considered, and it is argued that the symplectic form has finite behavior.  The solutions {\it do} have infinite values for their center of energies, suggesting their superselection.  Finally, the behavior of soft charges and their conservation is very briefly outlined for these solutions, which generalize the allowed values for soft charges.

\section{Non-antipodal solutions}

Most of the discussion of the present paper will be given for EM, although, based on previous work\cite{DoGi4,Gsplit2} much of this analysis is expected to have a straightforward gravitational extension.  The simplest example of a nonantipodal field configuration is that created by a Faraday line operator\cite{Dirac1955},
\beq\label{Farline}
e^{-iq\int_\Gamma A}\ ,
\eeq
where $A$ is the one form gauge potential, and $\Gamma$ is for example the positive $x$ axis, at $t=y=z=0$.
This operator can be used to dress a charge $q$ operator at $x^\mu=0$, making it gauge invariant, and 
creates at $t=0$ an electric field of the form
\beq
E^x = q \theta(x) \delta(y)\delta(z)\quad,\quad E^y=E^z=0\ 
\eeq
violating antipodal symmetry.

This field configuration is somewhat singular, and in particular has infinite energy.  To regulate the behavior of the energy at infinity, we can smear the field over a cone, by working in spherical polar coordinates $(r,\theta,\phi)=(r,\theta^A)$, defined with respect to the $x$ axis, and specifying an appropriate window function $f(\theta)$ with support localized near $\theta=0$:
\beq
E_f^r = \frac{f(\theta)}{r^2}\quad,\quad E_f^A=0\ .
\eeq
Here
\beq
 2\pi \int\sin\theta d\theta f = q
\eeq
to satisfy Gauss' law.  Since energy density is proportional to $\vec E^2$, this field has finite energy at infinity.  However, the singularity at the origin implies infinite energy there, and this is infinitely different from the energy of the Coulomb field $E_C$, which has $f=q/(4\pi)$.
To remedy that, we instead consider a field configuration of the following form:
\bea\label{matchfield}
E^i &=& E^i_C\quad,\quad r<R_1\ ,\cr
E^i &=& E_T^i\quad,\quad R_1<r<R_2\, \cr
E^i &=& E_f^i\quad,\quad R_2<r\ .
\eea
Here $E_T$ is a transitional field configuration, which smoothly interpolates between the Coulomb field at $r<R_1$ and the conical field at $r>R_2$, while satisfying Gauss' law; we can think of finding such a configuration by ``combing" field lines to smoothly match the two.  The field configuration \eqref{matchfield} has the same behavior as Coulomb, and in particular the same energy density, near the origin.  

If we consider initial conditions given by  \eqref{matchfield} and $B^i=0$, and assume the source charge stays fixed at $\vec x=0$, we expect this asymmetric field configuration to evolve towards the Coulomb field in the far future, together with outgoing radiation at ${\cal I}^+$.  Specifically, we can write such a  solution as the Coulomb field, plus a pure radiation field which has zero source.  We will focus on this radiation field $\cale$, with  initial conditions
\bea\label{radfield}
{\cal E}^i &=& 0\quad,\quad r<R_1\ ,\cr
{\cal E}^i &=& E_T^i -E_C^i\quad,\quad R_1<r<R_2\, \cr
{\cal E}^i &=& E_f^i-E_C^i\quad,\quad R_2<r\ .
\eea
Specifically, for $r>R_2$, ${\cal E}(t=0)$ has only a radial component
\beq\label{Er}
{\cal E}^r = \frac{f(\theta) - q/(4\pi)}{r^2} = \frac{g(\theta)}{r^2}\ .
\eeq
The preceding discussion implies that this radiation field has finite energy.

There have been numerous discussions of the role of antipodal symmetry, its importance for conservation laws, and the possibility that it is required by regularity conditions.  In view of the preceding construction, the latter in particular seems puzzling.  Specifically, with finite energy initial data, corresponding to an EM field that disperses to infinity, we might expect regular evolution.  This seems at odds with claims\cite{CaEy,HeTrEM,Prab,Esma,HeTrHD} of singular behavior at ${\cal I}^+$.  

These questions can be studied by finding the full evolution, with the initial data $E^i={\cal E}^i(\vec x)$, $\partial_t E^i=0$, and $B^i=0$ at $t=0$.  Note that Maxwell's equations then imply a nonzero initial value for $\partial_t B^i=-\nabla \times {\vec {\cal E}}$.  

Maxwell's equations also imply that the cartesian components of the electric and magnetic fields obey the scalar wave equation, $\Box \phi=0$. This means that the future evolution can be found from the retarded Green function,
\beq
G(x,x') = \frac{\delta(t-t'- |\vx'-\vx|)}{4\pi |\vx'-\vx|}\ ,
\eeq
satisfying $\Box' G(x,x') =- \delta^4(x-x')$.  Specifically, a Green's theorem argument then gives the electric field for $t>0$,
\beq\label{Esoln}
{\cal E} ^i(x) = -\partial _{t} \int d^3x'  \frac{\delta(t- |\vx'-\vx|)}{4\pi |\vx'-\vx|} {\cal E}^i(\vec x')\ ,
\eeq
where $\cale(\vec x) = \cale(0,\vec x)$.
We similarly find
\beq\label{Bsoln}
{\cal B}^i(x) = \epsilon^{ijk} \partial_j \int d^3x'  \frac{\delta(t- |\vx'-\vx|)}{4\pi |\vx'-\vx|} {\cal E}_k(\vec x')\ .
\eeq

These expressions are consistent with a gauge potential in radiation gauge, given by
\beq\label{Asoln}
A_i(x) =\frac{1}{4\pi t} \int d^3x'  \delta(t- |\vx'-\vx|) {\cal E}_i(\vec x')\quad,\quad A_0=0\ .
\eeq
The additional, Coulomb, condition $\nabla \cdot \vec A=0$ easily follows.
Given such explicit formulas for the solution of the EM field equations, the asymptotics are readily explored.

\section{Asymptotics}

For a given $x^\mu$, the delta functions in \eqref{Esoln}-\eqref{Asoln} restrict to integration in $\vx'$ over a sphere of radius $t$ about the point $\vx$.  
This may be parameterized by introducing $\vx''=\vx'-\vx$; the integrals are then over the sphere $|\vx''|=t$, and for example we find
\beq\label{Asimpl}
A_i(x) =\frac{t}{4\pi} \int d\Omega''   {\cal E}_i(\vx+t \hat x'')\ ,
\eeq
where $\hat x''$ is the unit vector in the $\vx''$ direction.

Null infinity ${\cal I}^+$ is approached by fixing $u=t-r$ and taking the limit $r\rightarrow\infty$.  Consider first the case $u<-R_2$, where the integrals \eqref{Esoln}-\eqref{Asimpl} only receive contributions from the region $r>R_2$.  Here, the cartesian components of $\cale(\vec x)$ are
\beq
\cale_x= \frac{g(\theta)}{r^2} \cos\theta\quad,\quad \cale_y=\frac{g(\theta)}{r^2} \sin\theta\cos\phi\quad,\quad \cale_z= \frac{g(\theta)}{r^2} \sin\theta\sin\phi\ .
\eeq
We then have bounds
\beq
|A_i(x)| \leq \frac{t\, {\rm Max} |g(\theta)|}{4\pi} \int d\Omega'' \frac{1}{(\vx+ t\hat x'')^2}\ .
\eeq
The latter integral may be performed by choosing the polar axis for $d\Omega''$ in the $\vx$ direction, giving
\beq
 \int d\Omega'' \frac{1}{(\vx+ t\hat x'')^2} = 2\pi \int_{-1}^1 \frac{d \cos\theta''}{r^2+2tr \cos \theta'' + t^2} = \frac{\pi}{rt}\int_{u^2}^{v^2} \frac{d\tau}{\tau} = \frac{\pi}{rt} \ln\left(\frac{v^2}{u^2}\right)\ ,
 \eeq
 where the substitution $\tau=(\vx+ t\hat x'')^2$ is used and $v=t+r$.  Consequently,
 \beq\label{abound}
|A_i(x)| \leq \frac{ {\rm Max} |g(\theta)|}{2r} \ln|v/u|\ ,
\eeq
for cartesian components $A_i$, which is finite at $\cali^+$.  Similar bounds hold for $\cale_i(x)$ and $\calb_i(x)$, for example
\beq\label{ebound}
|\cale_i(x)| \leq \frac{ {\rm Max} |g(\theta)|}{|uv|}\ .
\eeq
 
 These expressions do suggest a possible divergence at $u=0$.  While this would be potentially problematic if the radial form of $\cal E$ in \eqref{Er} held all the way to $r=0$, it does not.  For $u>-R_2$, the integrals with integrands given by \eqref{Er} are cut off at $r= R_2$.  This means that in the bounds \eqref{abound}, \eqref{ebound}, $u$ is replaced by $R_2$ for $u>-R_2$.\footnote{For $\cale_i$ there is an extra term from $r=R_2$, with similar asymptotics.}  There is also a contribution from the initial data in the region $R_1<r<R_2$.  However, that can also be bounded; for example the contribution to $A_i$ is bounded as $r\rightarrow\infty$ at fixed $u$ by
 \beq
 |A_i^{\rm ann}(x)| \leq \frac{ R_2^2\, {\rm Max}_{\rm annulus} |\cale_i|}{4t}\ .
 \eeq
 Thus there is no singular behavior at $u=0$.
 
 It is interesting to note the asymptotic behavior of the bounds \eqref{abound} and \eqref{ebound} as either $i^0$ or $\cali^+$ is approached.  In the first case, with $r\rightarrow\infty$ at fixed $t$, the bounds behave as
 \beq\label{ibound}
 |A_i| \lesssim { {\rm Max} |g(\theta)|} \frac{t}{r^2}\quad,\quad |\cale_i| \lesssim \frac{{\rm Max} |g(\theta)|}{r^2}
 \eeq
 and approaching $\cali^+$, with $r\rightarrow\infty$ at fixed $u$,
 \beq
 |A_i| \lesssim \frac{ {\rm Max} |g(\theta)|}{2r}\ln(2r/|u|)\quad,\quad |\cale_i| \lesssim \frac{{\rm Max} |g(\theta)|}{2r|u|}\ .
 \eeq
 
 It is also useful to examine the radial component $A_r$ more carefully, in preparation for studying soft charges.  In the region $u<-R_2$, expressions \eqref{Asimpl} and  \eqref{Er}, together with an expansion of $g(\theta)$ in Legendre polynomials, gives
\beq\label{Aexp}
A_r(\vx)= \frac{t}{4\pi} \int d\Omega''  \frac{\hat x'\cdot \hat x}{r^{\prime 2}} g(\theta) =  \sum_l g_l\, t\int \frac{d\Omega''}{4\pi}  \frac{\hat x'\cdot \hat x}{r^{\prime 2}}P_l(\hat x' \cdot \hat x_0)\ 
\eeq
 where $\hat x$ is the unit vector in the $\vx$ direction, and $\hat x_0$ denotes the original $x$ axis.  The terms in the expansion can be analyzed by defining $s=t/r$ and $\sigma = 1 + 2s \hat x\cdot \hat x'' + s^2$, and by using the the addition law for spherical harmonics
 \beq
 P_l(\hat x' \cdot \hat x_0) = \frac{4\pi}{2l+1} \sum_{m=-l}^{m=l} Y^*_{lm}(\hat x') Y_{lm}(\hat x_0)\ ,
 \eeq
 with angles defined with respect to the direction $\hat x$.  The individual terms in \eqref{Aexp} then become
 \beq
 t \int \frac{d\Omega''}{4\pi} \frac{\hat x'\cdot \hat x}{r^{\prime 2}}P_l(\hat x' \cdot \hat x_0) = \frac{t}{2r^2} \int_{-1}^1 d \cos \theta''\, \frac{\hat x'\cdot \hat x}{\sigma}\, P_l(\hat x'\cdot \hat x) \, \cdot \, P_l(\hat x_0\cdot \hat x)\ .
 \eeq
 Using $\hat x'\cdot \hat x= (\sigma + 1-s^2)/(2\sqrt\sigma)$ and changing integration variable to $\sigma$ then gives
 \beq
  t \int \frac{d\Omega''}{4\pi}  \frac{\hat x'\cdot \hat x}{r^{\prime 2}}P_l(\hat x' \cdot \hat x_0) = \frac{1}{4r} \int_{(1-s)^2}^{(1+s)^2} \frac{d\sigma}{\sigma} \frac{\sigma + 1-s^2}{2\sqrt\sigma} P_l\left(\frac{\sigma + 1-s^2}{2\sqrt\sigma}\right)  \, \cdot \, P_l(\hat x_0\cdot \hat x)\ .
  \eeq
  For even $l$, the integral over sigma produces a polynomial of $s$.  For odd $l$, the integral produces terms proportional to $(1-s^2)^k\ln\left(\frac{1+s}{1-s}\right)$, with $k$ ranging over integers up to $(l+1)/2$, plus polynomial terms.  Therefore, the series \eqref{Aexp} takes the form
\beq\label{Arasymp}
A_r(\vx)=\frac{1}{r}\sum_l g_l \left[(1-s^2)\ln\left(\frac{1+s}{1-s}\right)A_l(s) + B_l(s)\right]P_l(\hat x_0\cdot \hat x)\ ,
\eeq
where $A_l(s)$ are polynomials in $1-s^2$ which are nonvanishing only for $l$ odd, and $B_l(s)$ are polynomials of $s$.  

Eq.~\eqref{Arasymp} determines the asymptotics of $F_{tr} = \partial_t A_r$, which takes the form
\beq\label{Eexp}
F_{tr} = \frac{1}{r^2}\sum_l g_l \left\{2 \left[1- s \ln\left(\frac{1+s}{1-s}\right)\right]A_l(s) + (1-s^2)\ln\left(\frac{1+s}{1-s}\right)A_l'(s)  + B_l'(s)]\right\} P_l(\hat x_0\cdot \hat x)\ .
\eeq
The logarithmic behavior at odd $l$ was previously observed in \cite{HeTrEM}.  
In particular, as ${\cal I}^+$ is approached with $r\rightarrow\infty$ at fixed $u$, use of $s=1+u/r$ implies leading behavior 
\beq\label{Easymp}
F_{tr}\sim \frac{h(\theta)}{r^2} \ln(-u/r)\ .
\eeq
Notice that these expressions respect the asymptotic form of the bounds found in \eqref{abound} and \eqref{ebound}.

\section{Reexamination  of previous arguments}

Past work\cite{KPS,astrorev,CaEy,HeTrEM,Prab,Esma,SaWa,HeTrHD} has given various arguments for the importance of antipodal identification; these can be revisited, in light of the preceding discussion.

For example, arguments have been given\cite{HeTrEM,Prab,Esma,HeTrHD} that regularity on $\cali^+$ follows from antipodal symmetry.  The discussion of \cite{HeTrEM} appears in their appendix B.2,  where they  introduce the hyperbolic coordinates $\eta = \sqrt{|uv|}$, $s=t/r$ and consider the electric field component
\beq
E^\eta_{HT} = \frac{\eta^3}{(1-s^2)^2} F^{s\eta}=r^2 F^{tr}\ .
\eeq
With the asymptotics \eqref{Easymp}, this component indeed is singular as $r\rightarrow\infty$, as  observed in   \cite{HeTrEM,HeTrPC}.  

However, the usual electric field $F_{tr}$ is still  regular in this limit, in accord with the bounds in the preceding section; as was noted above, it would seem strange if it became singular for a finite energy configuration dispersing to infinity.

A closely related question is the behavior along the null cone $u=0$; since this also corresponds to $s=1$, the singular behavior in \eqref{Arasymp}-\eqref{Easymp} also suggests a finite-$r$ singularity there.  However, note that this is in the region $u>-R_2$, where the expressions \eqref{Arasymp}-\eqref{Easymp} no longer hold.  
One perspective on this modification comes from 
comparing to the analysis of \cite{HeTrEM}.  Maxwell's equations give their eq.~B.14, together with a term that did not appear with the asymptotics assumed there:
\beq\label{neweq}
\partial_s\left[(1-s^2)\partial_s E_{HT}^\eta\right] - D_A D^A E_{HT}^\eta - \frac{1}{\sqrt{ \gamma}} \partial_A \left(\sqrt{ \gamma} { \gamma}^{AB} \eta\partial_\eta F_{sB}\right)=0\ ,
\eeq
where $\gamma_{AB}$ is the round $S^2$ metric.
The  $u<-R_2$ asymptotics \eqref{abound} give large-$\eta$ scaling
\beq\label{etaasymp}
A_A \sim  \ln|v/u| = \ln\left(\frac{1+s}{1-s}\right)\quad ,\quad A_s = \frac{\eta s}{(1-s^2)^{3/2}} A_r \sim \frac{s}{1-s^2}\ln\left(\frac{1+s}{1-s}\right)
\eeq
and the extra term vanishes as in \cite{HeTrEM}; \eqref{neweq} then implies log singularities at $u=0$.  However, as was noted, for $u>-R_2$, this asymptotics is altered, replacing the logarithms in \eqref{etaasymp} with
\beq
\ln\left(\frac{t+r}{R_2}\right) = \ln\left(\frac{\eta}{R_2}\sqrt{\frac{1+s}{1-s}}\right)\ ;
\eeq
the $\ln \eta$ dependence, combined with the angular dependence, allows the extra term to contribute to \eqref{neweq}, and to the behavior of its solutions, invalidating the apparent argument for singular behavior in line with the physical expectations.

A second argument for antipodal symmetry has been based on finiteness of the symplectic structure.  For two solutions given by $\delta A_1$, $\delta A_2$, the (pre-) symplectic form is 
\beq
\Omega(\delta A_1, \delta A_2) = \int_\Sigma\left( \delta A_1\wedge \delta F_2 - \delta A_2\wedge \delta F_1\right)\ ,
\eeq
up to a possible term arising from gauge fixing, where $\Sigma$ is a Cauchy surface.  However, in the radiation gauge $A_0=0$, the radial field \eqref{Er} corresponds to $A_r=tg(\theta)/r^2$, and the combined expressions show that even the individual terms in the  form $\Omega$ are finite.

As pointed out in \cite{Gsplit2}, one kind of quantity {\it is} divergent for the generic configurations we have described; while the total energy and momenta are finite, the {\it boost charges}  $M_{0i}$ are divergent.\footnote{I thank M. Henneaux for the suggestion to check this.} This follows from the expression
\beq
M^{0i} = \int d^3x ( x^0 T^{0i} - x^i T^{00})
\eeq
for the boost charges, and the $t=0$ asymptotics \eqref{Er}.  The boost charge is physically interpreted as the ``center of energy," so it is not clear that there is a problem with this from a fundamental perspective, though it may mean that such configurations in effect correspond to different superselection sectors.  Of course, in any case behavior ``at infinity" involves describing an idealization of any physical configuration; for physical configurations of finite extent, we expect to have configurations that match the non-antipodal ones we have described arbitrarily well.  So, this question may just be one of how limits are defined.

\section{Soft charges and conservation laws}

Antipodal symmetry has received considerable emphasis in discussions of soft charges and conservation laws\cite{KPS,astrorev,HeTrEM}, and so it is interesting to investigate how that story changes in the presence of non-symmetric configurations.  This section will briefly outline some initial discussion of this question.

For an arbitrary function $\epsilon(\theta^A)$, the soft charges can be naturally defined at $i^0$ by \cite{HeTrEM}
\beq
Q_\epsilon^0 = \int_{i^0} \epsilon\, {}^*F\ .
\eeq
Soft charges can likewise be defined on constant $u$ or $v$ sections of $\cali^+$, $\cali^-$ as\cite{KPS}\cite{astrorev}
\beq\label{softscri}
Q^+_\epsilon(u) = \int_u  \epsilon\, {}^*F\quad,\quad Q^-_\epsilon(v) = \int_v  \epsilon\, {}^*F\ .
\eeq

Since $E^r$ is bounded by $1/r^2$ at $i^0$, the soft charges are well-defined there.  Antipodal symmetry would imply $Q_\epsilon^0=0$ for $\epsilon$ odd under parity, but the more general non-symmetric configurations have non-vanshing odd charges.  The configurations described in this paper thus exhibit a  generalization which allows non-zero values for all soft charges.

However, at $\cali^\pm$ the behavior \eqref{Easymp} implies that the expressions \eqref{softscri} in general diverge.   Despite this, it appears that {\it differences} in soft charges along $\cali^\pm$ are well-defined.  This follows from differentiating \eqref{Eexp}, which gives
\beq
\partial_u F_{tr} = \frac{1}{r^3}\sum_l g_l \left(-4\frac{s}{1-s^2} A_l(s)+\cdots\right)\ ,
\eeq
where subleading terms fall more rapidly with $r$.
The leading term of $r^2\partial_u F_{tr}$ is clearly finite (but nonzero) as $r\rightarrow\infty$ at fixed $u$.  
Thus, one might interpret the infinite part of the soft charges at $\cali^\pm$ in terms of an overall offset, which can be subtracted.  

Indeed, a general conservation law can be written for evolution along $\cali^+$ or $\cali^-$.  Considering the former, we have
\beq
\Delta Q= Q^+_\epsilon(u')-Q^+_\epsilon(u) = \int_u^{u'} d( \epsilon\, {}^*F) = \int_u^{u'} \left( d\epsilon\wedge {}^*F - \epsilon {}^*j\right)\ .
\eeq
Written in components, in Bondi coordinates $(u,r,\theta^A)$, this becomes
\beq
\Delta Q = -\int d\Omega du r^2  \left(\partial_A\epsilon\, F^{rA} + \epsilon j^r\right)\ ;
\eeq
note that $F^{rA}=\gamma^{AB}(F_{rB}-F_{uB})/r^2$.  This means that the soft charges evolve along $u$ either through electric current reaching $\cali^+$, or through tangential EM fields reaching $\cali^+$.  For example, the initially non-trivial soft charges $Q_\epsilon$ of a general non-antipodal configuration are expected to evolve to subtracted soft charges $Q_\epsilon^+(\infty)=0$, in the case with fixed electric charge source at the origin. 

Note that this picture is consistent with the discussion of \cite{DoGi4,Gsplit2},\footnote{For related discussion, see \cite{BoPo}.} where it is argued that the soft charges are characteristics of the EM field configuration, but not necessarily of the matter that serves as its source; for a given source, field configurations can be chosen with any or trivial soft charges, aside from the total electric charge, by addition of a general radiation field.

\section{Generalizations}

Previous investigation of allowed field configurations has focussed on those with antipodal symmetry\cite{KPS,astrorev,HeTrEM,HeTrHD}.  Since the present work has argued that there are regular configurations without antipodal symmetry, an interesting question is what is the full space of allowed boundary conditions, that results in finite energy, regular solutions.  As noted, the generic configurations considered in this paper do have infinite value for the center of energy, and so may be superselected.

 Antipodal symmetry has played a similar role in gravity.\footnote{See \cite{astrorev}, and its references.}  As was shown in \cite{DoGi1}, and further discussed in \cite{DoGi3,DoGi4,Gsplit2}, a natural class of diffeomorphism-invariant operators is that of  gravitational line operators, associated with choice of an axial or Fefferman-Graham-like gauge.  These break antipodal symmetry in a fashion directly analogous to the Faraday line operator \eqref{Farline}.  Considerations like in the rest of this paper are expected to extend.  A regulated version of these operators ({\it e.g.} smeared over a cone, and regulated at $r=0$) is expected to generically create a field that differs from the Coulomb (or linearized Schwarzschild) field by a finite-energy and momentum radiation field.  This field is thus expected to have regular evolution to $\cali^\pm$.  This indicates that the general class of regular, finite energy field configurations includes non-antipodally symmetric ones.  An interesting question is to characterize the general  boundary conditions that correspond to these.

\section{Acknowledgements}

I thank G. Compere and A. Zhiboedov for useful conversations, and particularly M. Henneaux and C. Troessaert for very valuable correspondence explaining their work.   I also thank the CERN theory group, where this work was carried out, for its hospitality.  This material is based upon work supported in part by the U.S. Department of Energy, Office of Science, under Award Number {DE-SC}0011702.

\bibliographystyle{utphys}
\bibliography{xantip}

\end{document}